\documentclass[useAMS,usenatbib]{mn2e}
\usepackage{latexsym,graphicx,natbib}
\usepackage{amsmath}

\newcommand\beq{\begin{equation}}
\newcommand\eeq{\end{equation}}

\newcommand\msun{\, \rm M_\odot}
\newcommand\pc{{\, \rm pc}}

\def\gsim{ \lower .75ex \hbox{$\sim$} \llap{\raise .27ex \hbox{$>$}} }
\def\lsim{ \lower .75ex\hbox{$\sim$} \llap{\raise .27ex \hbox{$<$}} }

\title[Eccentric binaries in rotating stellar systems]
{Massive black hole binary eccentricity in rotating stellar systems}

\author[Sesana et al.]  {Alberto Sesana$^{1}$, Alessia
  Gualandris$^{2}$ and Massimo Dotti$^{2,3}$\\ $^{1}$Albert Einstein
  Institut, Am M\"{u}hlenberg 1, Golm, D-14476,
  Germany.\\ $^{2}$Max-Planck Institut f\"{u}r Astrophysik,
  Karl-Schwarzschild-Str. 1, D-85741 Garching,
  Germany\\ $^{3}$Dipartimento di Fisica G.~Occhialini, Universit\`a
  degli Studi di Milano Bicocca, Piazza della Scienza 3, 20126 Milano,
  Italy }

\begin{document}

\maketitle

\begin{abstract}
In this letter we study the eccentricity evolution of a massive black
hole (MBH) binary (MBHB) embedded in a rotating stellar
cusp. Following the observation that stars on counter-rotating (with
respect to the MBHB) orbits extract angular momentum from the binary
more efficiently then their co-rotating counterparts, the eccentricity
evolution of the MBHB must depend on the degree of co-rotation
(counter-rotation) of the surrounding stellar distribution. Using an
hybrid scheme that couples numerical three-body scatterings to an
analytical formalism for the cusp-binary interaction, we verify this
hypothesis by evolving the MBHB in spherically symmetric cusps with
different fractions ${\cal F}$ of co-rotating stars.  Consistently
with previous works, binaries in isotropic cusps (${\cal F}=0.5$) tend
to increase their eccentricity, and when ${\cal F}$ approaches zero
(counter-rotating cusps) the eccentricity rapidly increases to almost
unity. Conversely, binaries in cusps with a significant degree of
co-rotation (${\cal F}>0.7$) tend to become less and less eccentric,
circularising quite quickly for ${\cal F}$ approaching unity. Direct
$N$-body integrations performed to test the theory, corroborate the
results of the hybrid scheme, at least at a qualitative level. We
discuss quantitative differences, ascribing their origin to the
oversimplified nature of the hybrid approach.

\end{abstract}


\section{Introduction}
Since their theoretical prediction in the early 80's \citep{br80},
massive black hole (MBH) binaries (MBHBs) forming in galactic nuclei
following galaxy mergers, have been the main focus of many dynamical
studies.  In the context of the hierarchical formation of galactic
structures \citep{wr78} and the MBHs residing at their center, bound
MBH binaries forming at parsec scales have to get rid of their orbital
energy to reach the point where gravitational waves (GW) emission
becomes efficient enough to drive their final coalescence.  This is
known as the 'last parsec problem' \citep{mm01}. Interactions with
ambient stars, abundant in dense nuclei, might provide the physical
source of energy extraction by means of the slingshot mechanism. That
is, in a strong three-body encounter, the light intruder star is
ejected at infinity carrying away part of the orbital energy and
angular momentum of the massive binary \citep{mv92}. In the last two
decades several analytical and numerical works \citep{qui96, mm01,
  hm02, mm03, mp04, mf04, bau06, ber06, ms06, mms07, mat07, shm08,
  ber09, pau10, s10} have been devoted to the study of MBHB dynamics in
galactic nuclei, the major focus being the evolution of the binary
semi-major axis to overcome the 'last parsec problem'. Most of these
works also report on the MBHB eccentricity evolution (which is in
general more difficult to track and much more affected by numerical
noise in $N$-body simulations), often observing a net increase during
the hardening process \citep[see][for a detailed
  discussion]{s10}. This is of particular importance because (i) GW
emission efficiency is a strong function of the eccentricity of the
system (with eccentric binaries coalescing much faster) and (ii) even
a small surviving eccentricity in the GW detection bands will be
easily detectable \citep{ps10}, possibly giving us clues about the
binary evolution.  Interestingly, the vast majority of the cited
papers \citep[with the notable exception of][]{ber06, pau10} considered 
the MBHB evolution in non-rotating stellar systems. There are,
however, three good reasons for considering rotating stellar distributions:

(i) observationally, classical galaxy bulges often show some degree of
net rotation \cite[see, e.g.][]{gadotti11}, and pseudobulges are
mainly rotationally supported \citep[e.g.][]{kormendy11}. A net
rotation is also observed in the central region the Milky Way
\citep{genzel96, schodel2009}.

(ii) MBHBs formed during galaxy mergers are embedded in the remnant of
the fusion of two galactic nuclei. Even assuming that the two nuclei
originally had no spin, the orbital angular momentum associated with
the merger will form a rotating system \citep{mm01}.

(iii) As noted by \cite{iwa10}, counter-rotating stars are much more
effective in extracting angular momentum from the MBHB.  We therefore
expect the MBHB eccentricity to evolve differently depending on the
degree of co(counter)-rotation of the surrounding stellar
distributions.
  
In this letter, we study the eccentricity evolution of a MBHB with a
small mass ratio (we consider $q \equiv M_2/M_1=1/81$) in stellar
cusps with different degrees of rotation (from purely co-rotating to
purely counter-rotating cusps). We apply the hybrid formalism
developed by Sesana et al. (2008, hereafter SHM08) that couples
numerical three-body scatterings to an analytical description of the
cusp-binary interaction. Given the several simplifying assumptions
adopted in the hybrid model that may lead to spurious results when it
comes to a delicate quantity like the binary eccentricity, we also
performed calibrated direct $N$-body integrations of the binary-cusp
system by means of the direct summation $N$-body code $\phi$GRAPE
\citep{Harfst2007}. The letter is organized as follows. In Section 2
we present a simple analytical argument to explain the different
behaviours of co-rotating and counter-rotating stars interacting with
the MBHBs, and we verify it with the support of calibrated three-body
experiments. In Section 3 we describe the hybrid and $N$-body models
used to integrate the joint MBHB-cusp evolution.  We present the
results of the two methods, discussing similarities and differences in
Section 4, and we conclude with some final remarks in Section 5.

\section{Analytical background and three body scattering}
In this section we discuss a simple argument to illustrate the
different behaviour of co-rotating and counter-rotating stars in the
star-binary interaction.  For the sake of simplicity, we focus on a
ideal coplanar case. A comprehensive analytical model will be
presented in a follow up paper. We consider a system consisting of:
(i) a binary with total mass $M=M_1+M_2$ ($M_2\ll M_1$), semi-major
axis $a$ and eccentricity $e$, with initial energy ${\cal
  E}=-GM_1M_2/(2a)$ and angular momentum ${\cal L}={\cal
  L}_{z}=\mu\sqrt{GMa(1-e^2)}$ (where $\mu=M_1M_2/M$) aligned along
the positive $z$ axis; (ii) a star either co-rotating or
counter-rotating with the binary , with $m_*\ll M_2$, in Keplerian
orbit around $M_1$ with semi-major axis $a_*\approx a$ and eccentricity
$e_*\approx e$. The star is characterized by an initial energy ${\cal
  E}_{*}\approx-GM_1m/(2a)$ and angular momentum ${\cal L}_{*}={\cal
  L}_{*,z}\approx \pm m_*\sqrt{GM_1a(1-e^2)}$ along the $z$ axis ($+$
if co-rotating with the binary, $-$ if counter-rotating).  Since
$M_2\ll M_1$, we ignore $M_2$ in the energy and angular momentum
budget of the stars, however such approximation (and the following
dissertation) works fairly well also in the case of mildly unequal
mass binaries with $q=M_2/M_1=1/3$.  In any case, in the following we
will consider $M_1\approx M$ (thus, $\mu \approx M_2$ ).  Setting
$a_*\approx a$ and $e_*\approx e$ is particularly convenient for
making a simple argument, since it cancels out complicated eccentricity
dependencies.
 
The starting point of our model is the definition of the MBHB
eccentricity as a function of its energy ${\cal E}$ and angular
momentum magnitude ${\cal L}$:
\begin{equation}
e=\sqrt{1-\frac{2{\cal E}{\cal L}^2}{GM^2\mu^3}}.
\label{eqecc}
\end{equation}
Differentiation of equation (\ref{eqecc}) leads to  
\begin{equation}
\Delta e=-\frac{(1-e^2)}{2e}\left(\frac{\Delta {\cal E}}{{\cal E}}+
\frac{2\Delta {\cal L}_z}{{\cal L}_z}\right)=\frac{(1-e^2)}{2e}\chi
\label{deltaecc}
\end{equation}
where $\chi$ is defined by the last equality.
We note that, being the binary angular momentum oriented along the $z$ axis, 
and being in general $\Delta {\cal L}\ll {\cal L}$, the eccentricity evolution 
depends on  $\Delta {\cal L}_z$ only. Exchanges 
of the $x$ and $y$ component of ${\cal L}$ will only result in a 
reorientation of the binary plane, without affecting $e$.
The sign of $\Delta e$ is therefore defined by the 
combination $\chi$ of the energy and angular momentum variations.

Following the three-body interaction, the star is ejected at infinity.
The ejection is usually caused by a close encounter with $M_2$, that
captures and ejects the star. This is known as the {\it slingshot
  mechanism}. In general, counter-rotating stars have larger relative
velocities with respect to $M_2$ then co-rotating ones, and therefore
their capture and ejection cross sections are much smaller.  In the
presence of a binary, stars experience an asymmetric potential that
allows for variations in the direction of their angular momenta.  As
shown in \cite{mgm09}, an eccentric binary exerts a semi-periodic
forcing in a direction perpendicular to the orbital plane of the
stars.  Since the torque acting on the orbital plane of the stars is
exerted by the binary, there is a correspondent change in the angular
momentum of the secondary, which in turn results in a change in the
eccentricity.  As a consequence of this torquing mechanism, both
initially co-rotating and counter-rotating stars undergo secular
evolution, and are ejected when they co-rotate with the binary
\citep{iwa10}.  Following this observation, we assume the star's final
energy to be negligible (i.e. ${\cal E}_{*_f}\approx 0$) and its
angular momentum $z$ component to be positive, of the form ${\cal
  L}_{*,z_f}=\eta m_*\sqrt{GM_1a(1-e^2)}$ (where in general
$0<\eta<1$).  For co-rotating stars we therefore have $\Delta{\cal
  L}_{*,z}=-\Delta{\cal L}_{z}=m_*(\eta-1)\sqrt{GM_1a(1-e^2)}$ and
$\Delta{\cal E}_{*}=-\Delta{\cal E}=GM_1m_*/(2a)$. Substituting in
Eq.~\ref{deltaecc} we obtain
\begin{equation}
\Delta e\propto\left(2\eta-3\right).
\label{decorot}
\end{equation}
The same reasoning applies to the counter-rotating stars, with the
exception that now the initial angular momentum is ${\cal
  L}_{*,z}=-m_*\sqrt{GM_1a(1-e^2)}$, leading to
\begin{equation}
\Delta e\propto\left(2\eta+1\right).
\end{equation}
This means that, unless $\eta>3/2$, the ejection of co-rotating stars decreases $e$
and, conversely, the ejection of counter-rotating stars increases $e$. 

\begin{figure}
  \begin{center}
    \includegraphics[width=8cm]{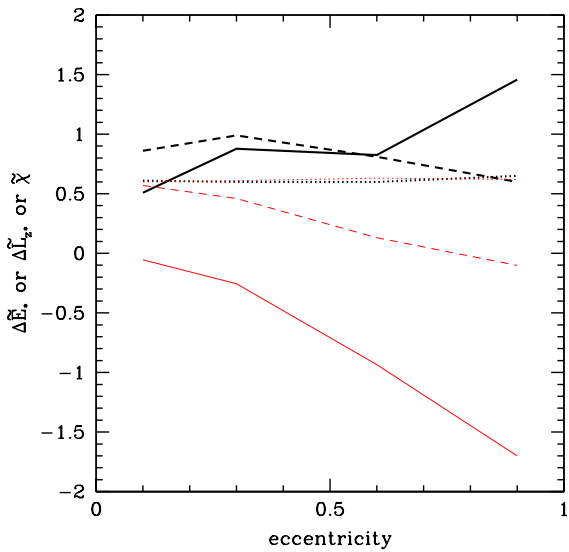}
  \end{center}
  \caption{Average star-binary exchanges in the three body interactions
    as a function of the MBHB eccentricity. Dashed curves:
    $\Delta\tilde{\cal L}_{*,z}=\Delta{\cal L}_{*,z}/{\cal L}_{*}^c$ 
    (i.e.  the $z$ component of the exchanged stellar
    angular momentum normalized to the angular momentum of a circular
    star with the same initial energy); dotted curves: ${\Delta\cal \tilde{E}}_{*}$
    (i.e. normalized to the binding energy of a star with $a_*=a$);
    solid curves $\tilde{\chi}=\chi(M_2/m_*)$. Thick black curves are
    for counter-rotating stars, thin red curves are for co-rotating
    stars.}
  \label{f3body}
\end{figure}

To test this simple heuristic description, we performed four sets of
three-body scattering experiments. We considered a binary with $q=1/81$
and different eccentricities $e=0.1$, 0.3, 0.6, 0.9. For each binary,
we integrated the orbit of 2000 stars with $a_*\approx a$,
eccentricity drawn from a thermal distribution $p(e)\propto e$ (to
mimic an isotropic distribution), and randomly oriented angular
momentum ${\cal L}_*$. We stored the energy and angular momentum of
each star after the ejection, dividing the stars in two groups: the
co-rotating (with ${\cal L}_{*,z}>0$) and the counter-rotating (with
${\cal L}_{*,z}<0$).  Figure~\ref{f3body} shows the numerical results
of the three-body experiments.  The average star-binary energy exchange
(dotted lines) is the same for both families and for any binary
eccentricity. The angular momentum exchange is instead always much
greater for counter-rotating stars, as expected. The practical
consequence of this is that the quantity $\chi \propto \Delta{e}$ is
positive for counter-rotating stars and negative for co-rotating stars
at any given binary eccentricity. Our simple argument fixes $e_*=e$,
to cancel out complications due to eccentricity, and therefore can not
catch the $\Delta{e}$ dependence on $e$. However, the main result of
our heuristic intuition is corroborated.

\section{Integration of the binary-star system}
Having understood the different average behaviour of co-rotating and
counter-rotating stars, we test here its practical consequences. We
follow the evolution of a MBHB in stellar cusps with different
degrees of rotation. Integrations are performed both using an hybrid
scheme and full direct $N$-body simulation. In the following, we briefly
describe the two techniques.

\subsection{The hybrid model}
SHM08 constructed an hybrid model for evolving unequal MBHBs in
stellar cusps. In short, it works as follows. They integrated
$5\times10^4$ stars in bound orbits around $M_1$. In the Newtonian
limit, the outcome of a scattering depends on $a_*/a \equiv x$, and
can be scaled to any absolute value of $a$.  Three-body scattering
experiments provide the distributions of the average star-binary
energy and angular momentum exchanges $\Delta {\cal E}(x)$ $\Delta
{\cal L}(x)$, together with the bivariate distribution of the ejection
times ${\cal N}^{\rm ej}(x,t)$. Such information, obtained
numerically, is then coupled to an analytical scheme for the joint
evolution of the binary and the surrounding stellar distribution. For
the practical integration of the hybrid model, a stellar
cusp with density $\rho(r)\propto r^{-\gamma}$, normalized at the
binary influence radius to an external isothermal sphere
($\rho(r)=\sigma^2/(2\pi G r^2)$), is assumed. The binary is placed with initial
eccentricity $e_i$ at an initial separation $a_i$ where the mass in
stars enclosed in its semi-major axis is twice the mass of $M_2$. In
the hybrid model, stars at each $x$ are weighted according to the
radial density distribution, a detailed description of the technique
is given in SHM08.

The averaging procedure washes out the different behavior of different
types of stars having similar values of $x$. However, here we want to
distinguish between co-rotating and counter-rotating stars. We
therefore build two sets of distributions $\Delta {\cal E}_{p/r}(x)$,
$\Delta {\cal L}_{p/r}(x)$, ${\cal N}^{\rm ej}_{p/r}(x,t)$, where $p$
(prograde) denotes the average over co-rotating stars only, and $r$
(retrograde) denotes the average over counter-rotating stars. Stellar
cusps with different degrees of rotation are constructed by mixing the
two distribution as ${\cal F}p+(1-{\cal F})r$, where ${\cal F}$ is a
parameter running from zero to one, describing the fraction of
co-rotating stars. We have ${\cal F}=0.5$ for an isotropic cusp while
fractions of ${\cal F}>0.5$ ($<0.5$) represent cusps with a net degree
of co-rotation (counter-rotation) with the binary.  Note that,
morphologically, the cusp is still spherically symmetric, i.e., this
approximation does not take into account any possible axisymmetry
or triaxiality induced in the cusp by the rotation. Moreover, being
based on three-body scattering experiments, precession effects due to
the extended potential of the cusp itself as well as other secular
collective effects are ignored.

\begin{figure*}
  \begin{center}
    \includegraphics[width=8cm]{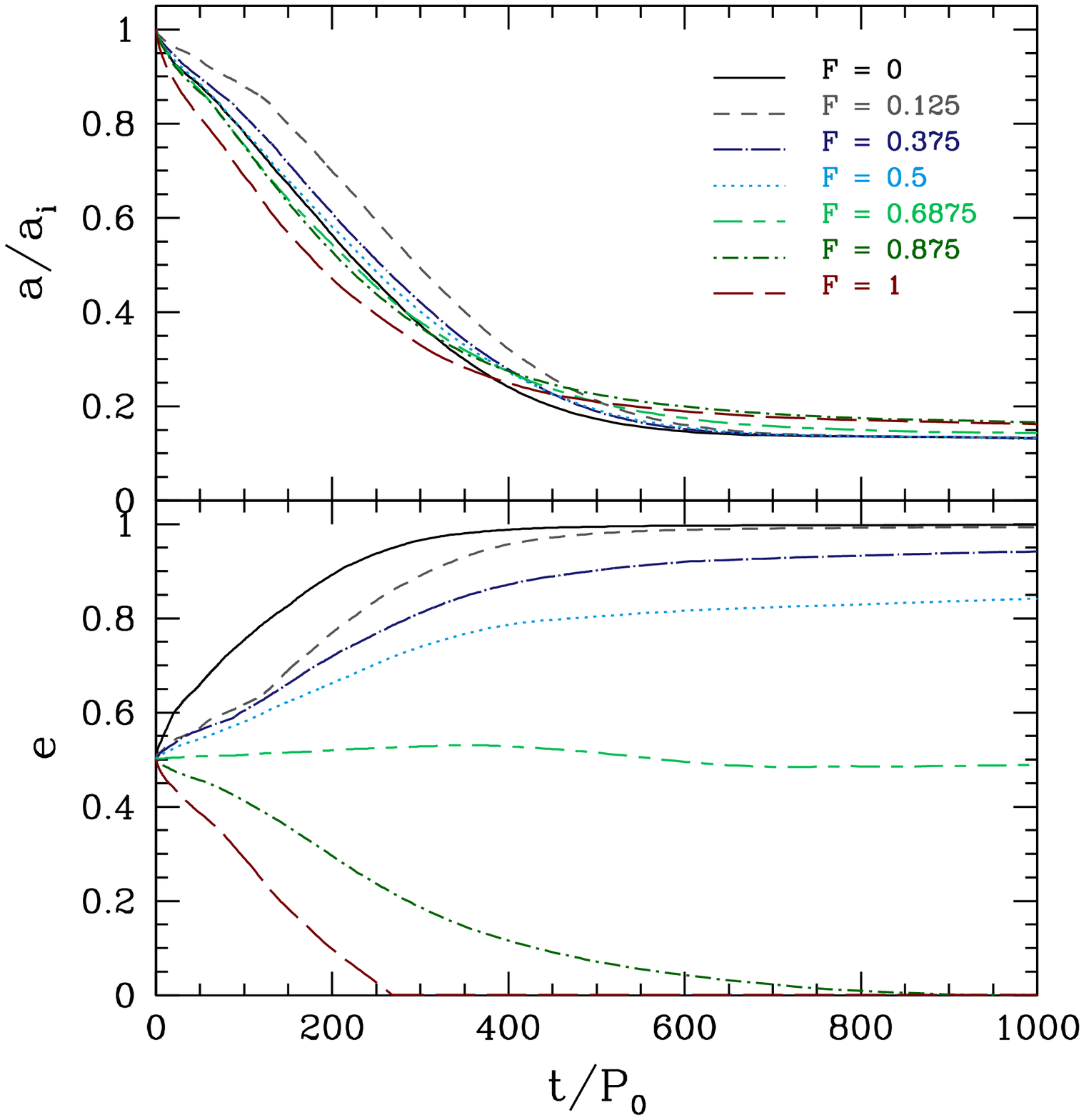}
    \includegraphics[width=8cm]{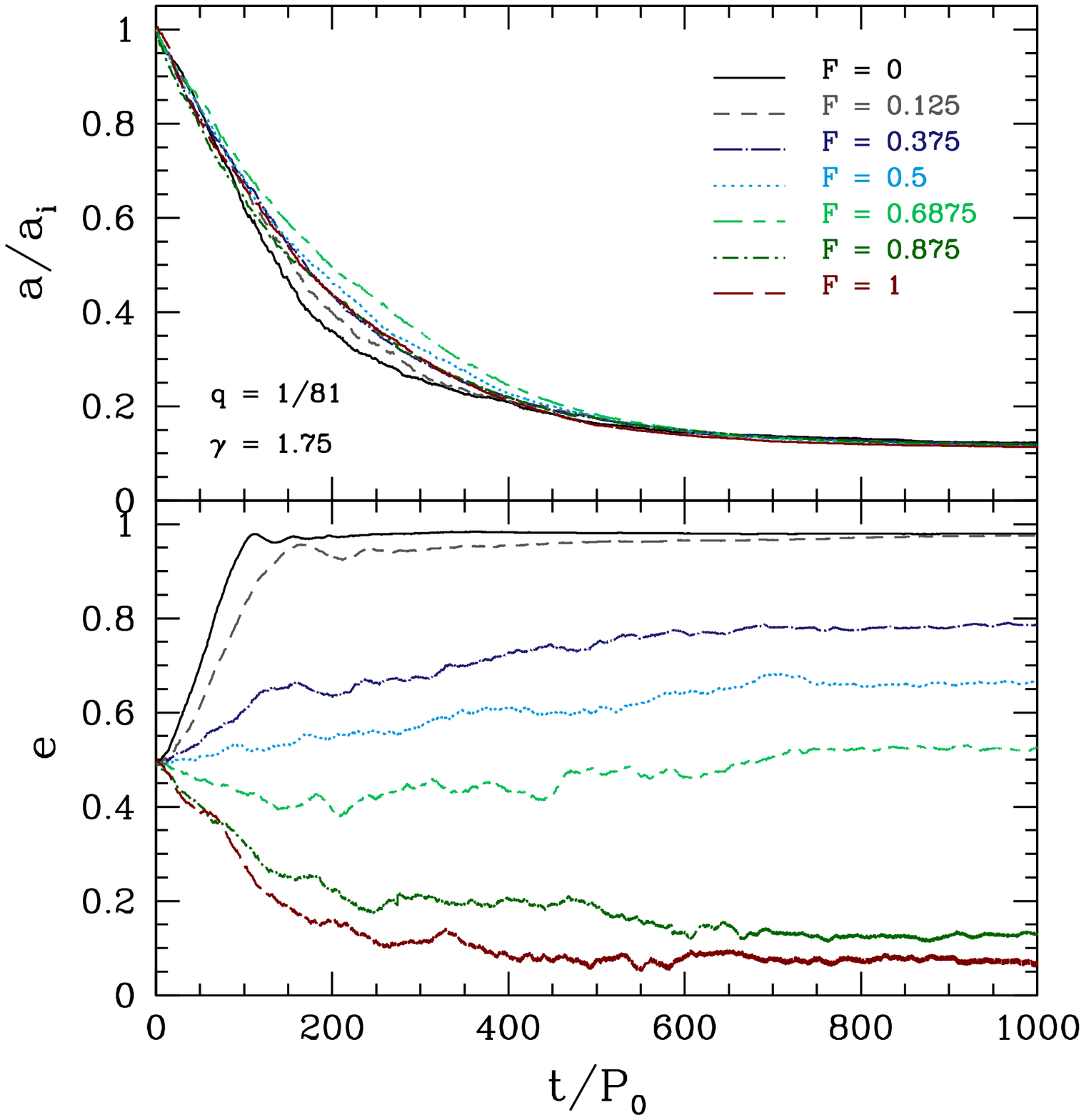}
  \end{center}
  \caption{{\it Left}: semi-major axis (upper panel) and eccentricity
    evolution (lower panel) of the MBHB according the the hybrid
    model. Different lines correspond to different fractions of
    co-rotating stars as labeled in the figure. We assume $q=1/81$,
    $\gamma=-7/4$ and $e_i=0.5$. $P_i$ is the Keplerian orbital
    period of the binary at $a_i$. {\it Right}: same for the direct
    $N$-body integrations. Line styles as in the left plot.}
  \label{figcompare}
\end{figure*}

\subsection{N-body simulations}
In order to test the hybrid model, we also performed $N$-body
simulations of a MBHB embedded in a power-law stellar
cusp and followed its evolution due to interactions with the stars.
All the runs were performed with the direct-summation parallel
$N$-body code $\phi$GRAPE \citep{Harfst2007} on the GPU enabled
computers at the Max-Planck Institute for Astrophysics in Garching.
The code uses a fourth-order Hermite scheme for the time-integration
and can be used in combination with GRAPE or GPU hardware, by means of
the {\tt Sapporo} library \citep{sapporo2009}.

While $N$-body simulations are computationally much more expensive
than three-body integrations and are therefore limited in particle
number, they allow to monitor the evolution of the black hole binary
under the combined effects of star interactions, Newtonian precession
and the Kozai mechanism. Precession of the orbits due to the
distributed stellar mass is not taken into account in the hybrid
model, but can play a role in the evolution of the binary. Moreover,
the cumulative effect of the individual scatterings can significantly
change the orientation of the binary orbital plane \citep{m02,gm07},
modifying the amount of cusp co-rotation as seen by the binary.

The binary-stellar distribution model was constructed to match the
initial set-up used in the hybrid scheme. 
We considered an unequal-mass binary with primary mass $M_1 =
10^6\msun$, mass ratio $q=1/81$, initial semi-major axis $a_i =
0.06\pc$ and initial eccentricity $e_i = 0.5$.  The stellar cusp follows a
Bahcall-Wolf $\rho(r) \sim r^{-7/4}$ profile at distances smaller than
$1\pc$, with total mass $M_{c} \sim 2.5\times 10^5 \msun$ and a mass 
enclosed in the binary orbit equal to $2 M_2$. Because the distribution 
function adopted for the generation of the initial conditions is approximated,
the constructed model is not in exact equilibrium. We therefore let
the cusp relax before adding the secondary black hole. During this
phase, the system undergoes a small expansion in the outer regions
which however does not affect the distance range where the secondary
black hole is placed. We use $N=32768$ for all models which
results in a black hole to star mass ratio of $ m_* / M_1 =
7.5\times10^{-6}$.

In addition to a model with an isotropic distribution of velocities,
we generate models with different fractions of co-rotating stars. These
are obtained by reversing the sign of all the velocity components for
a random subset of stars, at the time where the second black hole is added.
This procedure is effectively equivalent to the mixing procedure 
${\cal F}p+(1-{\cal F})r$ used in the hybrid scheme.

\section{Results}
The main results of our experiments are collected in
Fig.~\ref{figcompare}, where the MBHB semi-major axis and eccentricity
evolution in cusps with different ${\cal F}$ parameter is depicted. In
all runs we used $\gamma=-7/4$, $q=1/81$ and $e_i=0.5$. The temporal
evolution is plotted in units of the initial binary period $P_i$.  The
left plot in Fig.~\ref{figcompare} shows evolutionary tracks produced
by the hybrid integration scheme. The angular momentum of the star is
non influential in the energy exchange, hence the orbital decay of the
MBHB is hardly affected by the rotation of the cusp. The situation is
drastically different in the case of the eccentricity evolution (lower
panel). More counter-rotating cusps result in a faster evolution of
the binary toward higher eccentricities.  The critical fraction
defining the transition between eccentricity growth and
circularisation, for this particular choice of parameters, is ${\cal
  F}\approx0.7$.  Isotropic cusps (${\cal F}=0.5$) lead to significant
eccentricity growth, as found by SHM08. Results of direct $N$-body
integrations are shown in the right panels of Fig.~\ref{figcompare},
for the same values of ${\cal F}$ used for the hybrid model.  As
expected, we find that cusps with a larger fraction of stars on
co-rotating orbits with respect to the binary tend to circularise the
binary while cusps with a larger fraction of stars on counter-rotating
orbits tend to increase the binary eccentricity, in agreement with the
predictions from the hybrid model.  The timescale for the binary decay
is also in very good agreement with the model's results. There are,
however, some differences in the $N$-body results with respect to the
model. Firstly, the eccentricity in the $N$-body integrations does not
reach values as low and as high as in the hybrid model. In the fully
counter-rotating cusp (${\cal F}=0$), the eccentricity grows to $\sim
0.98$ while in the fully co-rotating cusp (${\cal F}=1$) the
eccentricity reaches $\sim 0.078$. This may be due to the small
particle number adopted in the simulations, but also to inaccuracies
in the hybrid scheme when interpolations are performed to the
eccentricity range boundaries. Also, the model with ${\cal F}=0.5$
shows only a very slow growth in the eccentricity, in contrast with
what is found in the hybrid scheme. This may be the result of the
suppression of the binary-induced secular evolution of the stars by
Newtonian precession related to the extended cusp. As a consequence,
counter-rotating stars are less prone to become co-rotating and are
less efficient in extracting angular momentum from the MBHB.  Lastly,
the eccentricity growth in the counter-rotating ${\cal F}=0$ case
appears much faster in the $N$-body results than in the hybrid
integration. This is due to the approximations used in the
semi-analytic prescription.  In the hybrid model, we subtract energy
and angular momentum to the binary at the moment of the star
ejection. Although this is a good approximation for the energy
transfer, it is not for the angular momentum.  Stars on
counter-rotating orbits secularly subtract angular momentum to the
binary to become co-rotating before being ejected. The MBHB
eccentricity growth therefore does not occur at the moment of the star
ejection (as in our hybrid scheme), but on a shorter timescale, during
the star-MBHB interaction. This is reflected in the much faster
eccentricity evolution in the counter-rotating N-body runs.

As discussed in Section 3.2, the cumulative torque exerted by the
stars changes also the orbital plane of the binary.  We find that the
orientation changes by $\approx 1-4$ degrees on the hardening
timescale ($a/\dot{a} \approx 100 P_0$) in all the runs but the two most
counter-rotating (${\cal F}=0, {\cal F}=0.125$).  This is in agreement
with the change predicted in Eq. 4 of \cite{gm07} for an unequal mass
binary. We note that, in the counter-rotating models, the large change
in the orbital plane (which results in the binary orbital angular
momentum reversal in the ${\cal F}=0$ case) occurs at $t/P_0>200$,
i.e. {\it after} the bulk of the orbital evolution. This is because
the large eccentricity attained by the binary (following the ejection
of a large number of counter-rotating stars) results in a very small
angular momentum. In this case, small torques can easily produce a
drastic reorientation of the binary plane. This dynamical aspect does
not affect our results, but is worth further investigation.

\section{Final remarks}

We demonstrate that the degree of rotation of the stellar background
surrounding a MBHB determines the eccentricity evolution of the
binary. This has already been discussed for wide BH pairs,
subject to dynamical friction exerted by a rotating background
\citep{dotti07}, and in the early evolution of pairing MBHBs in
rotating star clusters \citep{pau10}. This latter work, in particular,
describes the different action of dynamical friction in counter-rotating
clusters, and it is somewhat complementary to our paper. Here we 
highlight for the first time the physical mechanism at work 
in close binaries evolving through interactions with single stars. 
We find that, for an unequal mass
binary, stellar systems co-rotating with respect to the binary tend to
circularise its orbit. On the other hand, if stars have, on average,
angular momenta anti-aligned with respect to the binary, a strong increase
in the eccentricity is observed.
 
The dependence of the eccentricity evolution of the binary on the
rotation of the stellar background received so far little attention. 
Rotation can be due to secular evolution or merger events. For equal mass
binaries, presumably formed through major galaxy mergers, the MBHs are
likely to co-rotate with the nucleus of the remnant, reminiscent of the
orbital motion of the parent nuclei. In this case, more circular MBHBs
are expected. Very unequal mass binaries could form in situ
\citep[see][and references therein]{pau07}, or via galactic minor
mergers. In this case, the original rotation of the nucleus of the
primary galaxy would be less perturbed by the interaction, possibly
resulting in counter-rotating systems, and extremely eccentric
binaries.  In both equal and unequal cases, the exact eccentricity
evolution depends on the degree of rotation and on the slope of the
stellar profile. A more detailed analysis is postponed to a future
investigation.

\section*{Acknowledgments}
We thank Lucia Morganti and David Merritt for interesting discussions.


\begin{thebibliography}{}

\bibitem[Amaro-Seoane et al. (2007)]{pau07} Amaro-Seoane P., Gair
  J. R., Freitag M., Miller M. C., Mandel I., Cutler C. J. \& Babak
  S., 2007, Class. and Quantum Gravity, 24, 113

\bibitem[Amaro-Seoane et al. (2010)]{pau10} Amaro-Seoane P., Eichhorn C., Porter E. K. \& Spurzem, R., 2010, MNRAS, 401, 2268

\bibitem[Baumgardt, Gualandris \& Portegies Zwart (2006)]{bau06} Baumgardt H., Gualandris A. \& Portegies Zwart S., 2006, MNRAS, 372, 174 

\bibitem[Begelman, Blandford \& Rees (1980)]{br80} Begelman M. C., Blandford R. D. 
\& Rees M. J., 1980, Nature, 287, 307

\bibitem[Berczik et al. (2006)]{ber06} Berczik P., Merritt D., Spurzem R. \& Bischof H. P., 2006, ApJ, 642, 21

\bibitem[Berentzen et al. (2009)]{ber09} Berentzen I., Preto M., Berczik P., Merritt D. \& Spurzem R., 2009, ApJ, 695, 455

\bibitem[Dotti et al. (2007)]{dotti07} Dotti, M., Colpi M., Haardt F. \& Mayer L., 2007, MNRAS, 379, 956

\bibitem[{{Gaburov} {et~al.}(2009){Gaburov}, {Harfst}, \& {Portegies
  Zwart}}]{sapporo2009}
{Gaburov}, E., {Harfst}, S., \& {Portegies Zwart}, S. 2009, New A, 14, 630


\bibitem[{{Gadotti} (2011) {Gadotti}}] {gadotti11} Gadotti D.A., 2011, arXiv:1101.2714

\bibitem[{{Harfst} {et~al.}(2007){Harfst}, {Gualandris}, {Merritt}, {Spurzem},
  {Portegies Zwart}, \& {Berczik}}]{Harfst2007}
{Harfst}, S., {Gualandris}, A., {Merritt}, D., {Spurzem}, R., {Portegies
  Zwart}, S., \& {Berczik}, P. 2007, New Astronomy, 12, 357

\bibitem[Genzel et al. (1996)]{genzel96} Genzel R., Thatte N., Krabbe A., Kroker H. \& Tacconi-Garman L.E., 1996, ApJ, 472, 153

\bibitem[Gualandris \& Merritt (2007)]{gm07} Gualandris A. \& Merritt D., 2007, arXiv:0708.3038

\bibitem[Hemsendorf, Sigurdsson \& Spurzem (2002)]{hm02} Hemsendorf M., Sigurdsson S. \& Spurzem R., 2002, ApJ, 581, 1256

\bibitem[Iwasawa et al. (2010)]{iwa10} Iwasawa M., An S., Matsubayashi T., Funato Y. \& Makino J., 2010, arXiv:1011.4017


\bibitem[Kormendy, Bender \& Cornell (2001)] {kormendy11} Kormendy J.,
  Bender R. \& Cornell M.E., 2011, Nature, 469, 374


\bibitem[Makino \& Funato (2004)]{mf04} Makino J. \& Funato Y., 2004, ApJ, 602, 93

\bibitem[Matsubayashi, Makino \& Ebisuzaki (2007)]{mat07} Matsubayashi T., Makino J. \& Ebisuzaki T., 2007, ApJ, 656, 879

\bibitem[Merritt (2002)]{m02} Merritt D., 2002, ApJ, 568, 998

\bibitem[Merritt, Gualandris \& Mikkola (2009)]{mgm09} Merritt D., Gualandris A. \& Mikkola S., 2009, ApJ, 693, 35

\bibitem[Merritt, Mikkola \& Szell (2007)]{mms07} Merritt D., Mikkola S. \& Szell A., 2007, ApJ, 671, 53

\bibitem[Merritt \& Poon (2004)]{mp04} Merritt D. \& Poon M. Y., 2004, ApJ, 606, 788

\bibitem[Merritt \& Szell (2006)]{ms06} Merritt D. \& Szell A., 2006, ApJ, 648, 890

\bibitem[Milosavljevic \& Merritt (2001)]{mm01} Milosavljevic M. \& Merritt D., 2001, ApJ, 563, 34

\bibitem[Milosavljevic \& Merritt (2003)]{mm03} Milosavljevic M. \& Merritt D., 2001, ApJ, 596, 860

\bibitem[Mikkola \& Valtonen (1992)]{mv92} Mikkola S. \& Valtonen M.J., 1992, MNRAS, 259, 115

\bibitem[Perets \& Alexander (2008)]{pa08} Perets H. B. \& Alexander T., 2008, ApJ, 677, 146

\bibitem[Quinlan (1996)]{qui96} Quinlan G. D., 1996, NewA, 1, 35 

\bibitem[Porter \& Sesana (2010)]{ps10} Porter E. K. \& Sesana A., 2010, arXiv:1005.5296 

\bibitem[{{Sesana} {et~al.}(2008){Sesana}, {Haardt}, \& {Madau}}]{shm08}
{Sesana}, A., {Haardt}, F., \& {Madau}, P. 2008, ApJ, 686, 432

\bibitem[Sesana (2010)]{s10} Sesana A., 2010, ApJ, 719, 851

\bibitem[Sch\"odel, Merritt \& Eckart (2009)]{schodel2009} Sch\"odel R., Merritt D. \& Eckart 2009, A\&A, 502, 91


\bibitem[White \& Rees (1978)]{wr78} White S. D. M. \& Rees M. J., 1978, MNRAS, 310, 645


\end{thebibliography}
\end{document}